\documentclass[10pt,twocolumn,letterpaper]{article}

\usepackage{iccv}
\usepackage{times}
\usepackage{epsfig}
\usepackage{graphicx}
\usepackage{amsmath}
\usepackage{amssymb}

\usepackage{multirow}
\usepackage{tabularx}
\usepackage{booktabs}
\usepackage{amsfonts}
\usepackage{graphicx}
\usepackage{comment}
\usepackage{amsmath,amssymb} %
\usepackage{color}
\usepackage{enumerate}
\usepackage{cite}
\usepackage[export]{adjustbox} %
\usepackage{stmaryrd}   
\usepackage{makecell}
\usepackage{xcolor}
\usepackage{caption}
\usepackage{subcaption}

\usepackage[pagebackref=true,breaklinks=true,letterpaper=true,colorlinks,bookmarks=false]{hyperref}

\usepackage[capitalize]{cleveref}
\crefname{section}{Sec.}{Secs.}
\Crefname{section}{Section}{Sections}
\Crefname{table}{Table}{Tables}
\crefname{table}{Tab.}{Tabs.}

\newcommand{\topic}[1]
{
\vspace{1mm}\noindent\textbf{#1}
}

\newcommand{\MainMethodAbbr}{CEVR}
\newcommand{\AffineModuleHead}{Intensity transformation}
\newcommand{\AffineModule}{intensity transformation}

\newcommand{\continuousStackHead}{Continuous stack}
\newcommand{\continuousStack}{continuous stack}
\newcommand{\CycleTrainingHead}{Cycle training}
\newcommand{\CycleTraining}{cycle training}

\def\etal{et~al.}			  %
\def\eg{e.g.,~}               %

\iccvfinalcopy %

\def\iccvPaperID{6637} %

\begin{document}

\title{Learning Continuous Exposure Value Representations for\\Single-Image HDR Reconstruction}

\author{
Su-Kai Chen$^{1,2}$\quad
Hung-Lin Yen$^{1}$\quad
Yu-Lun Liu$^{1}$\quad
Min-Hung Chen$^{3}$\\
Hou-Ning Hu$^{2}$\quad
Wen-Hsiao Peng$^{1}$\quad
Yen-Yu Lin$^{1}$
\\
$^{1}$National Yang Ming Chiao Tung University \quad
$^{2}$MediaTek Inc. \quad
$^{3}$NVIDIA \quad\\
 {\url{https://skchen1993.github.io/CEVR_web/}}
}

\twocolumn[{%
\renewcommand\twocolumn[1][]{#1}%
\maketitle
\begin{center}
\centering
\captionsetup{type=figure}
\resizebox{0.90\textwidth}{!} 
{
    \includegraphics[width=\textwidth]{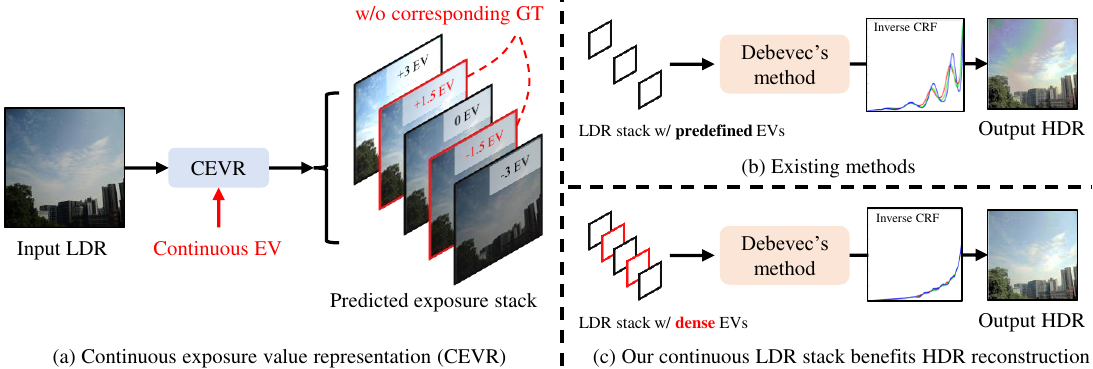}
    }
    
    \caption{\textbf{Single-image HDR reconstruction from continuous LDR stack.} 
    (a) Continuous Exposure Value Representation (CEVR) generates LDR images with continuous exposure values (EV) without corresponding ground truth during training.
    (b) Existing methods build LDR stacks only with EVs covered by training data, which brings less visible details for Debevec's method \cite{debevec1997recovering} to estimate an accurate inverse camera response function (CRF), resulting in artifacts on HDR results.
    (c) Our CEVR model enriches the LDR stack by including additional LDR images with continuous and dense EVs ({\color{red}{\textbf{red}}} frames), allowing Debevec's method to predict a more precise inverse CRF and reconstruct more visually pleasing HDR images.
    }
 	\label{teaser}
\end{center}
}]

\maketitle

\begin{abstract}
\vspace{-5mm}
Deep learning is commonly used to reconstruct HDR images from LDR images. LDR stack-based methods are used for single-image HDR reconstruction, generating an HDR image from a deep learning-generated LDR stack. However, current methods generate the stack with predetermined exposure values (EVs), which may limit the quality of HDR reconstruction. To address this, we propose the continuous exposure value representation ({\MainMethodAbbr}), which uses an implicit function to generate LDR images with arbitrary EVs, including those unseen during training. Our approach generates a \textit{\continuousStack} with more images containing diverse EVs, significantly improving HDR reconstruction. We use a \textit{\CycleTraining} strategy to supervise the model in generating continuous EV LDR images without corresponding ground truths. Our {\MainMethodAbbr} model outperforms existing methods, as demonstrated by experimental results.
\end{abstract}

\section{Introduction}
\label{sec:intro}

High dynamic range (HDR) images can capture detailed appearances in regions with extreme lighting conditions, like sun and shadow. As conventional cameras only capture a limited dynamic range in real-world scenes, one approach to address this issue is to blend multiple LDR images with different exposures into a single HDR image. However, this method is limited to static scenes and may result in ghosting or blurring artifacts in dynamic scenes. 
Additionally, this method is not applicable when multiple images of the same scene are unavailable, such as an image on the internet.

\begin{figure}[h!]
\centering
\resizebox{\columnwidth}{!}
    {%
    \begin{tabular}{ccccccc}
    \toprule
      \multirow{2}{*}{EV stack of \emph{real images}} &
      \multicolumn{2}{c}{\begin{tabular}[c]{@{}c@{}}PSNR\\ RH's TMO\end{tabular}} &
      \multicolumn{2}{c}{\begin{tabular}[c]{@{}c@{}}PSNR\\ KK's TMO\end{tabular}} &
      \multicolumn{2}{c}{HDR-VDP-2} \\ \cline{2-7} 
    & m     & $\sigma$ & m     & $\sigma$ & m     & $\sigma$ \\
    \midrule
    {[}-2, 0, 2{]}                       & 24.96 & 3.18     & 23.52 & 2.75     & 45.12 & 5.74     \\
    {[}-2, -1.3, 0, 1.3, 2{]}            & 25.15 & 2.98     & 23.71 & 3.12     & 45.26 & 5.12     \\
    {[}-2, -1.3, -0.7, 0, 0.7, 1.3, 2{]} & 25.32 & 3.01     & 23.92 & 3.04     & 45.51 & 4.86     \\ 
    \bottomrule
    \end{tabular}
    }
    \includegraphics[width=\columnwidth]{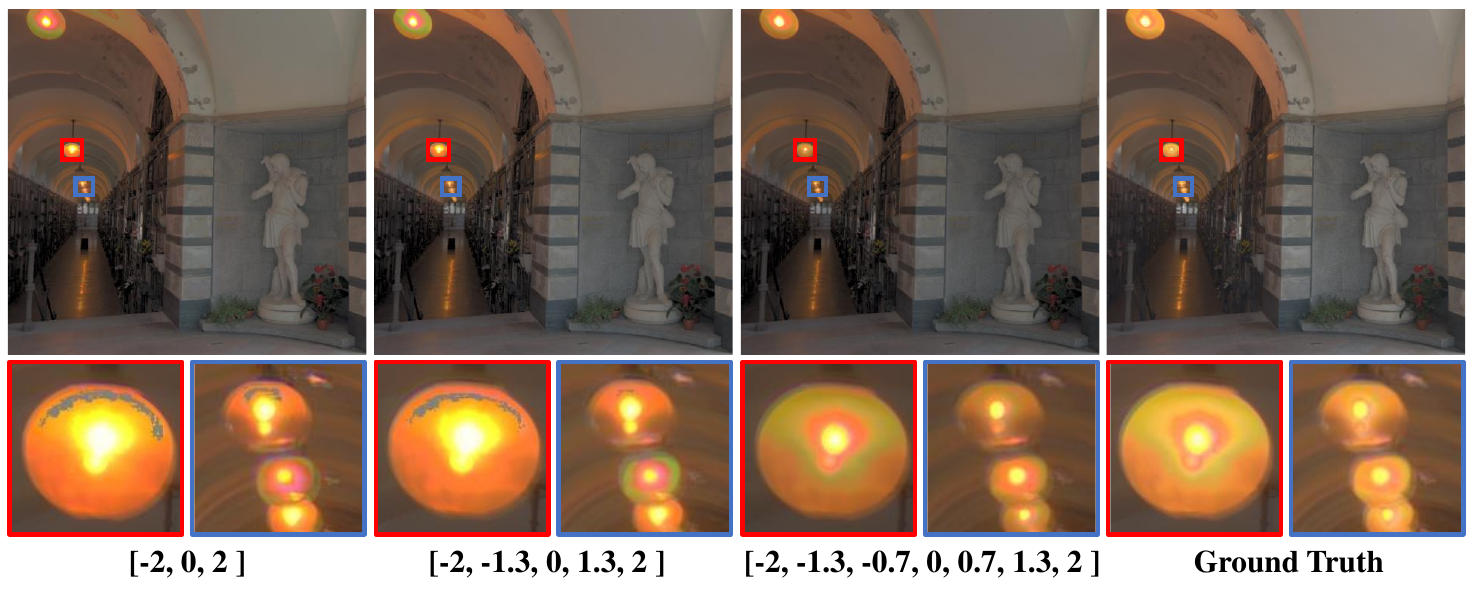}
    \caption{
            \textbf{Motivation.} 
            We observe that an LDR stack with dense EVs improves HDR reconstruction even with the same exposure range (from -2EV to +2EV). A list ``[-2,0,2]'' means the stack contains three LDR images with -2, 0, and 2 EVs. 
            An example of visual comparison is given.
    }
	\label{Motivation}
	\vspace{-3mm}
\end{figure}

Another branch of methods, e.g.,~\cite{eilertsen2017hdrcnn,endo2017drtmo,marnerides2018expandnet,Lee_2018,lee2018deep,santos2020single,liu2020single}, takes a single LDR image as input to generate the HDR counterpart without suffering from misalignment, which is referred to as  \textit{single-image HDR reconstruction}.
These approaches, e.g.,~\cite{Lee_2018,lee2018deep}, are trained on particular datasets and build an LDR stack with a single LDR image to generate an HDR image using Debevec's method~\cite{debevec1997recovering}.

Using more LDR images with richer EVs improves HDR image quality, as demonstrated in~\cref{Motivation} with different EV stack settings using Debevec’s method on the \emph{real} LDR images of the HDREye dataset~\cite{nemoto2015visual}. We compare tone-mapping operators RH~\cite{reinhard2002photographic} and KK~\cite{kim2008consistent} and use HDR-VDP-2 to evaluate HDR quality. However, accessible datasets have predefined and quantized EVs and may not cover optimal values for HDR reconstruction, causing information loss.

Previous studies \cite{chen2021learning,mildenhall2021nerf,sitzmann2020implicit} show the effectiveness of implicit neural representations in modeling continuous relationships, motivating our research.
Inspired by the observation in~\cref{Motivation}, we address the issue of predefined, quantized EVs by leveraging an implicit neural function to model relationships between image appearance and continuous EVs.
It turns out our method can generate LDR images with arbitrary EVs even if the corresponding ground truth is unavailable.
More importantly, LDR stacks enriched by images with these continuous and dense EVs can reconstruct HDR images of better quality.

Specifically, the proposed approach, \textit{continuous exposure value representation} (CEVR), exploits an implicit neural function to generate LDR images with continuous exposure values, as shown in ~\cref{teaser}(a).
Based on the flexibility of our CEVR model, we further develop two strategies, {~\CycleTraining} and {\continuousStack}, to improve the quality of the LDR stack and the final HDR result.

{~\CycleTrainingHead} utilizes CEVR to generate continuous EV images without relying on direct supervision from corresponding ground truths. We train the model using two continuous EVs that sum up to a predefined EV, with the proportion of these two continuous EVs randomly sampled. This strategy enforces the cycle consistency constraint, improving the model's ability to synthesize images with varying EVs and enhancing the quality of the LDR stack. We then use the enriched LDR stack containing seen and unseen EVs in training data for Debevec's method to produce more accurate inverse camera response functions (CRFs) and visually appealing tone-mapped images (\cref{teaser}(c)) compared to previous methods~\cite{Lee_2018,lee2018deep} (\cref{teaser}(b)).

Extensive evaluations demonstrate the effectiveness of our proposed {\continuousStack} and {\CycleTraining} on the VDS~\cite{Lee_2018} and HDREye~\cite{nemoto2015visual} datasets. Both quantitative and qualitative evaluations show that CEVR significantly outperforms existing methods.
The following summarizes our three primary contributions:
\begin{itemize}
\setlength\itemsep{0em}
\item We propose the CEVR approach, which can generate LDR images with continuous exposure values by modeling relationships between image appearances and exposure values.
\item With the flexibility of the CEVR model, we design a training strategy, {\CycleTraining}, to explore continuous EV information and enhance the quality of the estimated LDR stack.
\item We propose the {\continuousStack}, which consists of LDR images with continuous and dense exposure values and can improve the quality of final HDR images.
\end{itemize}

\section{Related Work}
\label{sec:RelatedWork}

\topic{Multi-image HDR reconstruction.}
Modern cameras typically have limited dynamic ranges and cannot well capture all visible details of a scene with a wide range of illumination.
To address this issue, one practical solution is to take multiple LDR images at different exposure levels and blend them into an HDR image.
To this end, conventional methods such as \cite{Picard95onbeing, debevec1997recovering} are developed to estimate the CRF~\cite{grossberg2003space}, upon which multiple LDR images are converted into the radiance field of the scene and transformed into an HDR image.
Recent methods, \eg \cite{kalantari2017hdr, wu2018deep}, use CNNs for directly fusing LDR images and reconstructing their HDR counterpart.
However, both conventional and CNN-based methods require multiple differently exposed images of a static scene.
Furthermore, for working on dynamic scenes, additional mechanisms are needed to alleviate the misalignment problem and avoid blurring or ghosting artifacts \cite{mangiat2010high, khan2006ghost, srikantha2012ghost}.
However, misalignment itself is a complicated issue to resolve.

\topic{Single-image HDR reconstruction.}
This task aims to reconstruct the HDR image using just one LDR input, also called inverse tone mapping~\cite{banterle2006itmo, banterle2009itmo, banterle2008itmo, banterle2007itmo}, and can bypass the misalignment problem. 
Existing methods need to enlarge the dynamic range \cite{akyuz2007hdr, rempel2007ldr2hdr, masia2017dynamic, serrano2016convolutional} and restore the lost details.
Generation techniques~\cite{goodfellow2014generative, zhang2018perceptual, yu2018free, yu2018generative, liu2018image} for image synthesis are essential to methods of this category.
Due to the superior mapping power of CNNs \cite{he2016deep,VGG} and GAN~\cite{goodfellow2014generative}, deep neural networks are widely adopted for HDR reconstruction.

One branch of research efforts~\cite{eilertsen2017hdrcnn,marnerides2018expandnet, yang-cvpr18-DRHT,santos2020single, zhang2017learning} focuses on learning the mapping from the input LDR image to the HDR image.
For example, Marnerides et al. \cite{marnerides2018expandnet} use CNNs to generate the HDR image based on an LDR input.
To further improve the performance, Santos et al.~\cite{santos2020single} filter out the saturated regions in the LDR input and pretrain the deep network for an inpainting task.
However, learning the LDR-to-HDR mapping is ill-posed since  different LDR images can be mapped to the same HDR image \cite{endo2017drtmo}.

Another branch of methods, e.g., \cite{endo2017drtmo,Lee_2018,lee2018deep,kim2021end}, aims to synthesize a stack of differently exposed LDR counterparts given an LDR image as input.
Then, the conventional multi-image methods can be applied to the synthesized LDR stack to complete HDR reconstruction.
For example, Endo et al. \cite{endo2017drtmo} use 3D convolutions, with exposure variation being one dimension, to learn the relationship between the LDR input and its counterparts with different exposure values.
Their approach can generate the LDR stack directly.
The LDR stack can be synthesized in a recursive manner~\cite{Lee_2018,lee2018deep,kim2021end}.
For example, Lee et al. \cite{lee2018deep} use GAN to generate an image with relative exposure value change. 
The LDR stack is constructed by recursively using their model.

Nevertheless, existing stack-based methods can only generate LDR images with predefined exposure values present in the training data.
Inspired by the fact that the real-world captured images can have any EV value depending on different shutter settings instead of predefined ones, we present a method that can synthesize LDR images with continuous exposure values that are even unseen in the training data.
Our method can generate an enriched and denser stack with which significantly better HDR results are achieved.

\topic{Implicit neural representations.}
An implicit function space is a shared function space that contains the neural representation of different objects or images learned by a shared implicit function. It is commonly a latent space where a latent code is mapped to an image using an encoder-decoder structure~\cite{chen2019learning,mescheder2019occupancy, saito2019pifu, saito2020pifuhd, xu2019disn}. This approach has been widely used in image super-resolution~\cite{chen2021learning, lee2022local}, 3D shape, surface modeling~\cite{chen2019learning, Atzmon_2020_CVPR, sitzmann2020implicit}, and view synthesis of 3D structures~\cite{mildenhall2021nerf, niemeyer2020differentiable}. Methods using implicit functions have shown that the learned latent space can be continuous~\cite{Park_2019_CVPR, mildenhall2021nerf, chen2021learning, chibane2020implicit}, allowing for exploring continuous relationships of exposure differences between images.

More and more radiance field reconstruction research aims to generalize the trained model across scenes unseen in training data.
The methods in \cite{wang2021ibrnet, chen2021mvsnerf, yu2021pixelnerf} propose advanced model architectures and training strategies, making the learned implicit function space achieve the generalization on unseen views. Our method, similar to \cite{wang2021ibrnet, chen2021mvsnerf, yu2021pixelnerf}, can generalize well to all images without fine-tuning.

\begin{figure*}[t]
    \begin{subfigure}{\textwidth}
    \includegraphics[width=1\textwidth]{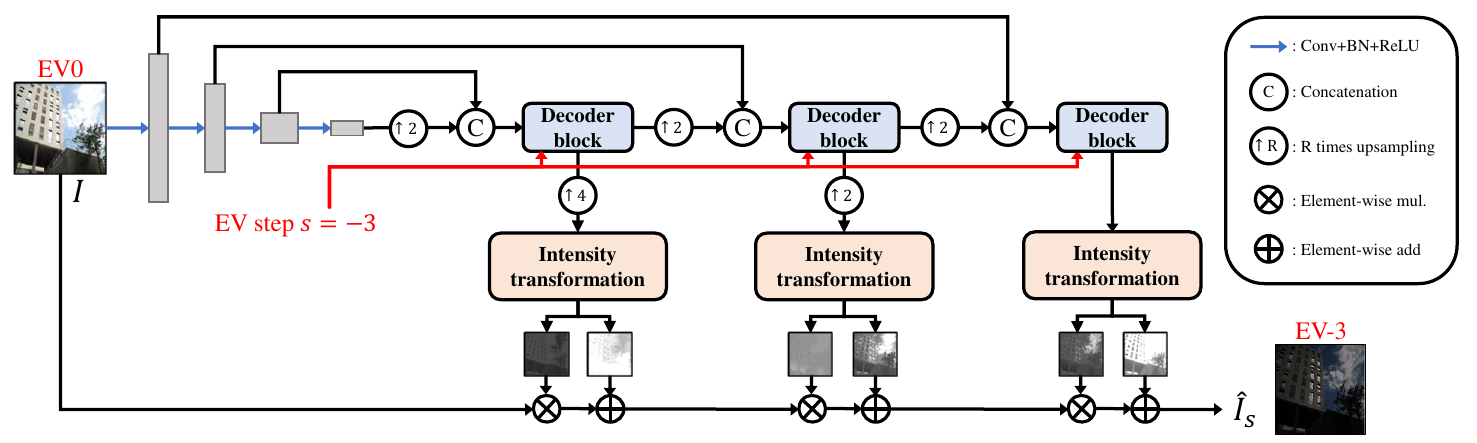}
    \vspace{-0.1in}
    \caption{CEVR network}
    \end{subfigure}%
    
    \medskip %
    \begin{subfigure}{0.31\textwidth}
    \includegraphics[width=1\textwidth]{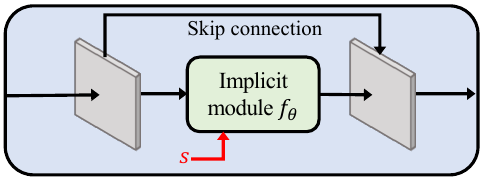}
    \vspace{-0.1in}
    \caption{Decoder block}
    \end{subfigure}%
    \hfill
    \begin{subfigure}{0.31\textwidth}
    \includegraphics[width=1\textwidth]{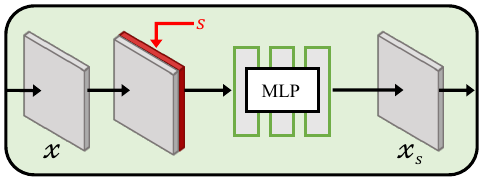}
    \vspace{-0.1in}
    \caption{Implicit module}
    \end{subfigure}
    \hfill  
    \begin{subfigure}{0.31\textwidth}
    \includegraphics[width=1\textwidth]{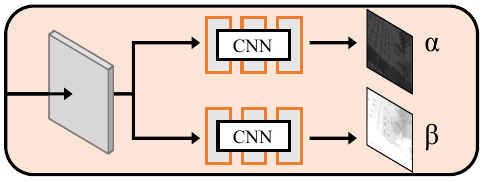}
    \vspace{-0.1in}
    \caption{Intensity transformation}
    \end{subfigure}%
    \vspace{-0.05in}
    \caption{
            \textbf{Proposed network architecture.} 
            (a) The proposed CEVR model takes an image $I$ and an EV step $\textit{s}$ as input, and produces an LDR image $\hat{I}_\textit{s}$ with a relative exposure value change \textit{s}.
            It adopts the U-Net structure, where the encoder is a pre-trained VGG-Net, and the decoder is a cascade of decoder blocks.
            (b) Each decoder block comprises an implicit module to enable continuous EV representation learning, as shown in (c).
            (d) Following each decoder block, an intensity transformation module is learned to produce the $\alpha$ and $\beta$ maps for image brightness transformation.            
    }
    \label{Method_archi}
    \vspace{-0.1in}
\end{figure*}

\section{Approach}
\label{sec:Approach}
In this section, we present our proposed Continuous Exposure Value Representation (CEVR) which generates LDR images with continuous EV. 
We provide an overview of our method in Section \ref{sec:Overview}, followed by the architectural design in Section \ref{sec:CEVR_structure}, which includes the implicit module and \AffineModule. 
Additionally, we propose two strategies, {\CycleTraining}, and {\continuousStack}, to further enhance the flexibility of CEVR, which are discussed in detail in Section \ref{sec:Cycle_training} and \ref{sec:Continuous_Stack}, respectively.

\subsection{Overview}
\label{sec:Overview}

Based on the observation in~\cref{Motivation}, we propose the CEVR model to generate an enriched and denser LDR stack for high-quality HDR reconstruction.
Our model, shown in~\cref{Method_archi}, utilizes a hierarchical U-Net structure (\cref{Method_archi}(a)) and incorporates the implicit neural representation into the design to predict LDR images with continuous EVs (\cref{Method_archi}(c)). 
To maintain accurate color and image structure while adjusting brightness, we introduce {\AffineModule} (\cref{Method_archi}(d)), which generates an adjustment map from each scale of the feature map.
As the ground-truth LDR images with unseen exposure values are lacking, we train the model using unsupervised {\CycleTraining} (\cref{Cycle_training}), enabling our method to learn images with varying EVs and enhance the quality of the predicted LDR stack.

\subsection{Continuous Exposure Value Representation}
\label{sec:CEVR_structure}
We show our CEVR model in~\cref{Method_archi}(a). 
Our CEVR model employs the hierarchical U-Net structure, where the encoder is a pre-trained VGG-Net and the decoder is a cascade of decoder blocks (\cref{Method_archi}(b)), each of which comprises an implicit module that compiles the feature map with an input EV step $\textit{s}$. Each decoder block is followed by an {\AffineModule} to adjust the intensity of input image at that scale.
Specifically, the CEVR model $F$ takes an LDR image $I$ and the specified EV step $\textit{s}$ as input and generates another LDR image $\hat{I}_\textit{s}$, a counterpart of $I$ with the relative exposure value change $\textit{s}$, via
\begin{equation}
\hat{I}_\textit{s} =F(I, \textit{s}).
\label{CEVR_eq}
\end{equation}

Take the widely used VDS dataset~\cite{Lee_2018} as an example. An LDR image with EV0 in this dataset can serve as $I$. 
An LDR stack can be generated by applying our CEVR $F$ to $I$ and every EV step in $\{\textit{s} \in \mathbb{Z}|-3 \leq \textit{s} \leq 3\}$. 

\topic{Implicit module.}
To synthesize an LDR image conditioned on a continuous EV step $\textit{s}$ even unseen in the training data, each decoder block in~\cref{Method_archi}(b) has an associated, learnable implicit module $f_{\theta}$, which is built by MLPs and shown in~\cref{Method_archi}(c). The implicit module $f_{\theta}$ parameterized by $\theta$ takes the form:
\begin{equation}
x_\textit{s}(p, q) = f_{\theta}([x(p, q), \textit{s}]), 
\label{ImplicitModule_eq}
\end{equation}
where $x\in\mathbb{R}^{H \times W \times C}$ is the input feature map, \(x(p,q) \in \mathbb{R}^{C}\) is the feature vector at location $(p,q)$, and \([x(p,q), s] \in \mathbb{R}^{C+1} \) refers to the concatenation of $x(p,q)$ and $s$.
The output feature map $x_\textit{s}$ is generated by repeatedly applying the implicit module $f_{\theta}$ to all $H \times W$ locations of $x$ with the desired relative exposure value change $\textit{s}$.

\topic{Intensity transformation.} 
In~\cref{Method_archi}(a), our CEVR leverages U-Net to perform multi-scale synthesis to generate a better LDR image with a different EV.
The input and output images, $I$ and $\hat{I}_\textit{s}$, cover the same scene under different exposures.
Thus, their content should not undergo significant changes.
To preserve the image structure and allow the model to focus on the brightness changes for detail reconstruction at each scale, the proposed intensity transformation module in \cref{Method_archi}(d) takes the resized feature map from the decoder block as input and produces the $\alpha$ and $\beta$ maps.
As shown in~\cref{Method_archi}(a), the $\alpha$ and $\beta$ maps carry out affine brightness transformation at each scale.
The output $\hat{I}_\textit{s}$ is synthesized through multi-scale transformations.

\topic{Reconstruction Loss.}
Suppose that we are given a training set of $N$ images $\{I_n\}_{n=1}^N$ with a set of $M$ EV steps $\{\textit{s}_m\}_{m=1}^M$. 
For each training image $I_n$, its ground-truth LDR stack $\{I_n^*(\textit{s}_m)\}_{m=1}^M$ is provided, where $I_n^*(\textit{s}_m)$ is the counterpart of $I_n$ with the relative exposure value change $\textit{s}_m$.
We train the CEVR model $F$ in Eq. (\ref{CEVR_eq}) by minimizing the $L_1$ reconstruction loss:
\begin{equation}
\mathcal{L}_{\text{rec}} = \sum_{n=1}^N \sum_{m=1}^M  \| I_n^*(\textit{s}_m) - F(I_n, \textit{s}_m) \|_1.
\label{eq:rec}
\end{equation}

\subsection{Cycle Training Strategy}
\label{sec:Cycle_training}

\begin{figure}[]
	\centering
	\includegraphics[width=\columnwidth]{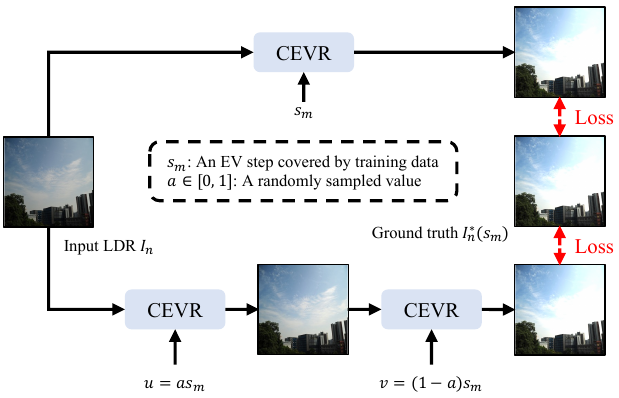}
	\vspace{-5mm}
    \caption{\textbf{Cycle training.} We derive the CEVR model in an unsupervised {\CycleTraining} strategy without using the corresponding ground truth. In this way, our model exploits the cycle consistency constraint and learns more continuous information by varying the EV sub-step $\textit{u}$.
    }
	\label{Cycle_training}
	\vspace{-2mm}
\end{figure}

Existing training datasets, such as the VDS dataset \cite{Lee_2018}, provide the ground truth for a sparse set of predefined EV steps, \eg $[-3, -2, ..., 3]$.
Inspired by the success of cycle consistency training in video frame interpolation~\cite{liu2019deep} and to make our CEVR work well for synthesizing images with arbitrary EVs, we introduce the {\CycleTraining} strategy to train the model with \emph{continuous} EV steps without the corresponding ground-truth images.
For each training image $I_n$ and each EV step $\textit{s}_m$ covered by the training set, the {\CycleTraining} strategy shown in~\cref{Cycle_training}, derives the CEVR model with two branches.
The first branch takes $I_n$ and $\textit{s}_m$ as input.
Since the ground-truth image $I_n^*(\textit{s}_m)$ is available, the reconstruction loss $\mathcal{L}_{\text{rec}}$ is used to supervise this branch.

The second branch implements a two-step process.
We randomly sample a real value $a \in [0,1]$ for each image $I_n$ at each training iteration, and decompose the EV step $\textit{s}_m$ into two sub-steps: $\textit{u} = a\textit{s}_m$ and $\textit{v} = (1 - a)\textit{s}_m$, with $\textit{u}+\textit{v}=\textit{s}_m$.
Our CEVR model is applied \emph{twice} with the two EV sub-steps, respectively.
Although the ground truth for the randomly sampled sub-step $\textit{u}$ is unavailable, we expect that the output of taking the two sub-steps should be similar to the ground truth $I_n^*(s_m)$ because of $\textit{u}+\textit{v}=\textit{s}_m$. Thereby, we enforce the proposed cycle loss:
\begin{equation}
\mathcal{L}_{\text{cyc}} =  \sum_{n=1}^N \sum_{m=1}^M \| I_n^*(\textit{s}_m) - F(F(I_n, \textit{u}), \textit{v}) \|_1.
\label{eq:cycle}
\end{equation}

The sub-step $\textit{u}$ in Eq. (\ref{eq:cycle}) is randomly sampled for each training image with each covered EV step at each training iteration.
It is used to simulate arbitrary EV step input to our CEVR model.
To compensate for the lack of the ground truth of the intermediate output $F(I_n, \textit{u})$, the cycle loss $\mathcal{L}_{\text{cyc}}$ in Eq. (\ref{eq:cycle}) offers indirect supervision, ensuring the continuity of our CEVR model with continuous EV steps.

The objective function used to derive the proposed CEVR is defined by
\begin{equation}
\mathcal{L} = \mathcal{L}_{\text{rec}} + \lambda \mathcal{L}_{\text{cyc}},
\end{equation}
where we empirically set \(\lambda\) to 0.1 in our experiments.

\subsection{Continuous Stack}
\label{sec:Continuous_Stack}
In the inference phase, with the implicit module and the {\CycleTraining} strategy, our CEVR model can generate high-quality LDR images with continuous EVs. 
The LDR stack containing more LDR images with various EVs can help Debevec’s method~\cite{debevec1997recovering} estimate a more accurate inverse CRF, as shown in~\cref{teaser}(c), and improve the HDR image reconstruction, as shown in \cref{Motivation}. 
Inspired by this observation, we proposed the {\continuousStack}, which predicts additional LDR images with continuous EVs from our CEVR model.
The predicted continuous and dense LDR stack further benefits the stack fusion process and enhances the final HDR quality, as shown in~\cref{teaser}(c).

\section{Experiments}
\label{sec:Experiment}
\subsection{Experimental Setup}

\topic{Datasets.} 
We train our model using the training set of the VDS dataset~\cite{Lee_2018}, which contains image stacks of 48 scenes.
The testing sets of the VDS and HDREye datasets~\cite{nemoto2015visual}, which contain 48 and 42 scenes respectively, serve as the testing sets for evaluations.
The auto-bracketing feature of the camera produces seven photos with predefined exposure values for each scene in the VDS dataset (EV-3 to EV+3). We follow the common evaluation protocol~\cite{Lee_2018,lee2018deep} and select the image with the zero exposure value, which is expected to have the most evenly distributed histogram, as the input to the model.

\topic{Training details.}
For training, we consider each training scene $n$ from the VDS dataset~\cite{Lee_2018} and take the corresponding EV0 LDR image as input $I_n$.
We also take each EV step $\textit{s}_m \in \{-3, -2, -1, 0, 1, 2, 3\}$ into account.
We feed $I_n$ and $\textit{s}_m$ to the CEVR model and estimate the LDR image with EV $\textit{s}_m$ for model training.
Since the inverse CRF is usually asymmetrical, we train two different models with the same architecture to handle the increasing and decreasing exposure changes, respectively. 
For upsampling, we use bicubic upsampling, followed by a $3\times3$ 2D convolution with stride 1 and padding 1.
The model is trained for 1,250 epochs with Adam optimizer \cite{kingma2014adam} and cosine annealing warmup with restarts as the scheduler.
We use random rotation and flip to augment the data.
\begin{table}[t!]
    \caption{\textbf{Quantitative comparison of the predicted LDR stacks on the VDS dataset~\cite{Lee_2018}.}
    CEVR outperforms existing approaches in estimating LDR stacks for all EVs. With {\CycleTraining}, our method can generate high-quality LDR images even with large EV changes. 
    }
    \centering
    \resizebox{\columnwidth}{!}{%
        \begin{tabular}{clcccccc}
            \toprule
                \multirow{2}{*}{EV} & \multirow{2}{*}{Method} & \multicolumn{2}{c}{PSNR} & \multicolumn{2}{c}{SSIM} & \multicolumn{2}{c}{MS-SSIM} \\ \cline{3-8} 
                                      &                & m     & $\sigma$     & m     & $\sigma$      & m     & $\sigma$     \\ 
                                      \midrule
                \multirow{3}{*}{+3}   &  Deep chain HDRI\cite{Lee_2018}         & 28.18 & 2.77 & 0.953 & 0.065 & 0.983 & 0.015 \\
                                      &  Deep recursive HDRI\cite{lee2018deep}  & 28.97 & 2.92 & 0.944 & 0.044 & 0.981 & 0.014 \\
                                      &  {\MainMethodAbbr} (Ours)                        & \textbf{34.34} & 3.46 & \textbf{0.973} & 0.021 & \textbf{0.989} & 0.007 \\
                                      \midrule
                \multirow{3}{*}{+2}   &  Deep chain HDRI\cite{Lee_2018}         & 29.65 & 3.06 & 0.959 & 0.065 & 0.986 & 0.016 \\ 
                                      &  Deep recursive HDRI\cite{lee2018deep}  & 29.43 & 2.85 & 0.952 & 0.039 & 0.986 & 0.010 \\
                                      &  {\MainMethodAbbr} (Ours)                          & \textbf{35.30} & 3.08 & \textbf{0.981} & 0.016 & \textbf{0.993} & 0.004 \\
                                      \midrule
                \multirow{3}{*}{+1}   &  Deep chain HDRI\cite{Lee_2018}         & 31.90 & 3.43 & 0.969 & 0.039 & 0.992 & 0.008 \\
                                      &  Deep recursive HDRI\cite{lee2018deep}  & 32.02 & 2.85 & 0.969 & 0.026 & 0.992 & 0.006 \\
                                      &  {\MainMethodAbbr} (Ours)                          & \textbf{37.64} & 2.96 & \textbf{0.989} & 0.009 & \textbf{0.996} & 0.004 \\
                                      \midrule
                \multirow{3}{*}{-1}   &  Deep chain HDRI \cite{Lee_2018}        & 29.01 & 3.83 & 0.935 & 0.056 & 0.980 & 0.017 \\ 
                                      &  Deep recursive HDRI\cite{lee2018deep}  & 31.22 & 3.69 & 0.951 & 0.031 & 0.986 & 0.090  \\
                                      &  {\MainMethodAbbr} (Ours)                       & \textbf{34.62} & 3.47 & \textbf{0.980} & 0.011 & \textbf{0.992} & 0.005 \\
                                      \midrule
                \multirow{3}{*}{-2}   &  Deep chain HDRI\cite{Lee_2018}         & 26.72 & 4.54 & 0.952 & 0.029 & 0.974 & 0.021 \\
                                      &  Deep recursive HDRI\cite{lee2018deep}  & 31.08 & 3.07 & 0.948 & 0.041 & 0.986 & 0.014 \\
                                      &  {\MainMethodAbbr} (Ours)                        & \textbf{33.89} & 4.34 & \textbf{0.978} & 0.017& \textbf{0.988} & 0.010 \\
                                      \midrule
                \multirow{3}{*}{-3}   &  Deep chain HDRI\cite{Lee_2018}         & 24.33 & 4.57 & 0.919 & 0.036 & 0.948 & 0.037 \\
                                      &  Deep recursive HDRI\cite{lee2018deep}  & 29.15 & 4.75 & 0.910 & 0.061 & 0.966 & 0.025 \\
                                      &  {\MainMethodAbbr} (Ours)                       & \textbf{30.58} & 5.32 & \textbf{0.954} & 0.046 & \textbf{0.972} & 0.032 \\
                                      \bottomrule
        \end{tabular}
    }
    \vspace{1mm}    
    \label{LDR_stack}
    \vspace{-4mm}
\end{table}

\begin{figure}[t!]
	\centering
	    \includegraphics[width=1.0\columnwidth]{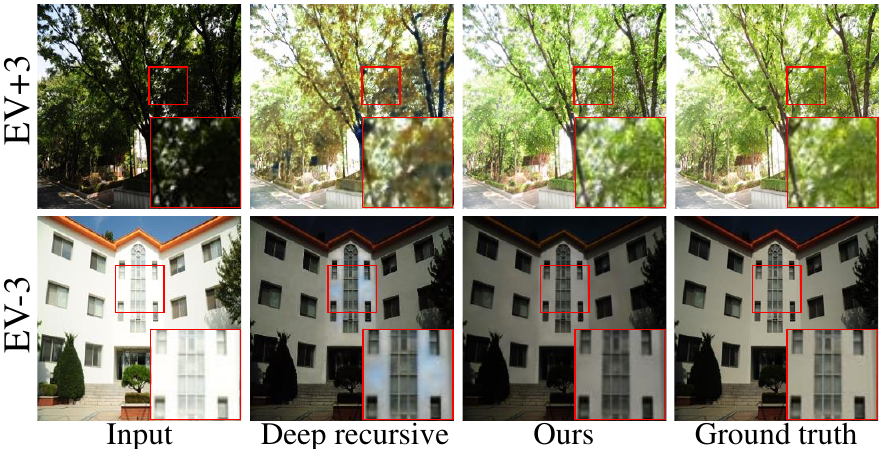}
	\vspace{-7mm}
	\caption{
        \textbf{Qualitative comparison of LDR image predictions in the VDS dataset~\cite{Lee_2018}.} 
        Our approach recovers more details compared to Deep recursive HDRI~\cite{lee2018deep} in the EV-3 example. In the EV+3 example, our approach generates LDR images with a color tone similar to the ground truth.
        }
	\label{LDR_stack_detail}
	\vspace{-5mm}
\end{figure}

\topic{Evaluation metrics.} 
We employ PSNR, SSIM~\cite{wang2004image}, and MS-SSIM~\cite{wang2003multiscale} as the metrics for evaluating the qualities of the predicted LDR stacks and HDR tone-mapped images.
We also utilize HDR-VDP-2~\cite{mantiuk2011hdr}, a metric based on the human visual system, to evaluate the quality of the reconstructed HDR images. We follow the setting of~\cite{Lee_2018,lee2018deep}, which sets a 24-inch monitor with a viewing distance of 0.5 meters, a peak contrast of 0.0025, and a gamma of 2.2 for measuring the HDR-VDP-2 metric.

\begin{figure*}[t!]
	\centering
	    \includegraphics[width=0.95\linewidth]{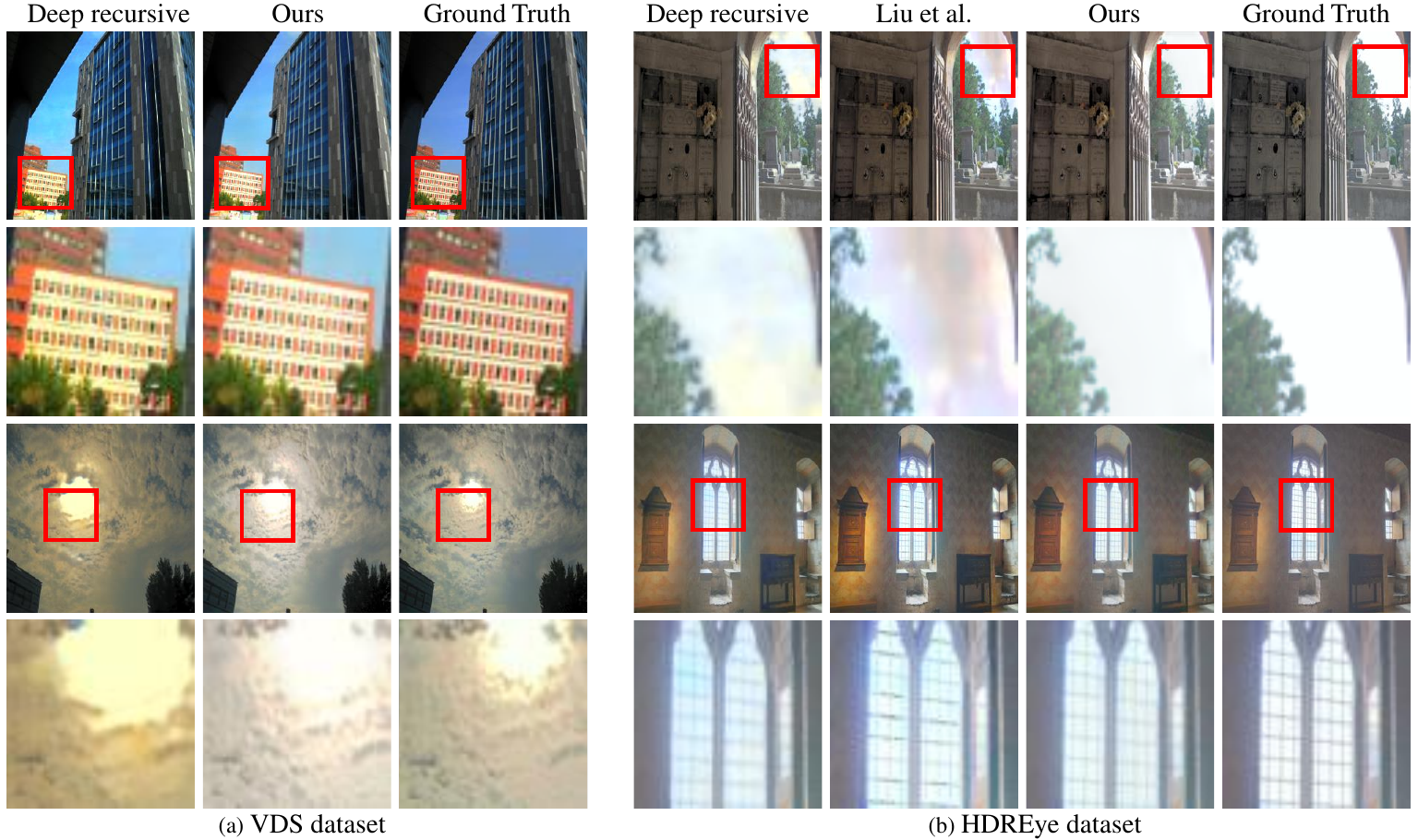}
	    \vspace{-2mm}
	\caption{
        \textbf{Qualitative comparison of tone-mapped HDR images.} 
        We adopt Reinhard's method~\cite{reinhard2002photographic} to generate HDR images shown in the figure. 
        Deep recursive HDRI\cite{lee2018deep} often suffers from erroneous color tones and artifacts.
        As shown in the figure, our method can better recover details when compared to Liu et al.'s method \cite{liu2020single}.
        In most regions, our method reconstructs the tone-mapped images with more accurate high-frequency details, leading to visually pleasing results.
        }
	\label{HDR_TMO_img}
	    \vspace{-2mm}
\end{figure*}

\topic{HDR reconstruction and tone-mapping operators.}
Our approach uses Debevec's approach~\cite{debevec1997recovering} to reconstruct HDR images with the predicted LDR stack and utilizes Reinhard's method~\cite{reinhard2002photographic} or Kim and Kautz's method~\cite{kim2008consistent} to tone-map the HDR images.

\subsection{Comparison of LDR Stacks Prediction}

\topic{Quantitative comparisons.}
The quantitative comparisons of the estimated LDR exposure stacks from the VDS dataset are shown in \cref{LDR_stack}. 
The table shows that the proposed method performs favorably against existing methods at every exposure value.
The output LDR image quality decreases as the exposure value gap increases because more extensive over- and under-exposed regions reconstruction are required, which makes the task more difficult.
However, with our {\CycleTraining}, our model can still generate high-quality LDR images in the cases of EV+3 and EV-3 by incorporating continuous and dense EV information into the training process. The continuous EV generation during training helps the model learn how to explicitly infer LDR images with arbitrary exposure values. 

\topic{Qualitative comparisons.}
With the {\CycleTraining}, our method can generate a high-quality LDR image even with large EV changes.
A detailed qualitative comparison is presented in \cref{LDR_stack_detail}. The first row of this figure shows that Deep recursive HDRI~\cite{lee2018deep} often gets less accurate color tone in estimating the LDR images, which may further degrade the quality of the HDR images fused by the LDR stack.
On the contrary, our method can better estimate the LDR images in all EVs with more accurate color tones. 
In addition, the second row demonstrates that our method can estimate the LDR images without severe artifacts. More visual comparisons can be found in the supplementary material.

\begin{table}[t]
\caption{\textbf{Quantitative comparison of HDR and TMO images.} Two tone-mapping approaches, Reinhard's approach~\cite{reinhard2002photographic}, and Kim and Kautz's approach~\cite{kim2008consistent}, are denoted as RH's and KK's TMO.}
\centering
\resizebox{\columnwidth}{!}{%
    \begin{tabular}{clcccccc}
    \toprule
      \multirow{2}{*}{Dataset} &  \multirow{2}{*}{Method} & \multicolumn{2}{c}{\begin{tabular}[c]{@{}c@{}}PSNR\\ RH's TMO\end{tabular}} &
      \multicolumn{2}{c}{\begin{tabular}[c]{@{}c@{}}PSNR\\ KK's TMO\end{tabular}} &
      \multicolumn{2}{c}{HDR-VDP-2} \\\cline{3-8}
                            &\multicolumn{1}{c}{}                  &m       &$\sigma$ &m       &$\sigma$ &m       &$\sigma$  \\
                            \midrule
    \multirow{6}{*}{VDS}    &DrTMO\cite{endo2017drtmo}             & 25.49  & 4.28    & 21.36  & 4.50    & 54.33  & 6.27 \\
                            &Deep chain HDRI\cite{Lee_2018}        & 30.86  & 3.36    & 24.54  & 4.01    & 56.36  & 4.41 \\
                            &Deep recursive HDRI\cite{lee2018deep} & 32.99  & 2.81    & 28.02  & 3.50    & 57.15  & 4.35 \\
                            &Santos et al. \cite{santos2020single} & 22.56  & 2.68    & 18.23  & 3.53    & 53.51  & 4.76 \\
                            &Liu et al. \cite{liu2020single}       & 30.89  & 3.27    & 28.00  & 4.11    & 56.97  & 6.15 \\
                            &{\MainMethodAbbr} (Ours)              & \textbf{34.67}   & 3.50    & \textbf{30.04}  & 4.45    & \textbf{59.00}  & 5.78 \\
                            \midrule
    \multirow{6}{*}{HDREye} &DrTMO\cite{endo2017drtmo}             & 23.68  & 3.27    & 19.97  & 4.11    & 46.67  & 5.81 \\
                            &Deep chain HDRI\cite{Lee_2018}        & 25.77  & 2.44    & 22.62  & 3.39    & 49.80  & 5.97 \\
                            &Deep recursive HDRI\cite{lee2018deep} & 26.28  & 2.70    & 24.26  & 2.90    & 52.63  & 4.84 \\
                            &Santos et al. \cite{santos2020single} & 19.89  & 2.46    & 19.00  & 3.06    & 49.97  & 5.44 \\
                            &Liu et al. \cite{liu2020single}       & 26.25  & 3.08    & 24.67  & 3.54    & 50.33  & 6.67 \\
                            &{\MainMethodAbbr} (Ours)         & \textbf{26.54}  & 3.10    & \textbf{24.81}  & 2.91    & \textbf{53.15}  & 4.91 \\
                            \bottomrule
                            
    \end{tabular}
}
    \vspace{1mm}
\label{HDR_TMO}
	\vspace{-2mm}
\end{table}

\subsection{Comparison of HDR Image Prediction}
We compare our method with five recent single-image HDR reconstruction methods, including Santos~\etal~\cite{santos2020single}, DrTMO~\cite{endo2017drtmo}, Deep chain HDRI~\cite{Lee_2018}, Deep recursive HDRI~\cite{lee2018deep}, and Liu~\etal~\cite{liu2020single}.
For Santos~\etal~\cite{santos2020single}, Deep recursive HDRI~\cite{lee2018deep} and Liu~\etal~\cite{liu2020single}, we use their official implementations along with the released pre-trained model weight to generate all the quantitative and qualitative results on the VDS~\cite{Lee_2018} and HDREye~\cite{nemoto2015visual} datasets. For DrTMO~\cite{endo2017drtmo} and Deep chain HDRI~\cite{Lee_2018}, we compare our results to the numbers reported in their papers.
For HDR image prediction, our approach adopts the {\continuousStack} strategy where the EV steps are enriched from $\{-3, -2, ..., +3\}$ to $\{-3, -2.5, ..., +3\}$, and the images with the extra EV steps are also synthesized by using the proposed CEVR model.

\topic{Quantitative evaluations.}
As shown in \cref{HDR_TMO}, our method performs favorably against the competing methods on the VDS dataset~\cite{Lee_2018}.
The HDREye dataset~\cite{nemoto2015visual} serves as a blind test bed, our HDR prediction still achieves better qualities using the same tone-mapping operators. 
Our proposed {\CycleTraining} makes the model explicitly learn the continuity as EV steps change, leading to better generalization on the unseen dataset, HDREye. With the \continuousStack, more LDR images with various EVs are involved in the fusion process, which helps Debevec’s approach \cite{debevec1997recovering} estimate a more accurate inverse CRF and generate HDR images with better qualities.

\topic{Qualitative comparisons.}
To generate the tone-mapped images for visual comparisons, we first reconstruct HDR images by fusing the LDR stacks with Debevec's approach~\cite{debevec1997recovering}. 
Then we use Reinhard's method~\cite{reinhard2002photographic} to generate HDR TMO images for all comepting methods. 
As Liu et al. \cite{liu2020single} generate an HDR image directly, we apply Reinhard's tone-mapping operator~\cite{reinhard2002photographic} for tone-mapping the HDR images.
Note that Liu et al.~\cite{liu2020single} do not train their method on the VDS dataset; hence we only compare the qualitative results with their method on the HDREye dataset, as shown in~\cref{HDR_TMO_img}.

In \cref{HDR_TMO_img}, it can be observed that Deep recursive HDRI~\cite{lee2018deep} often suffers from inaccurate color tone: the color of the building is inaccurate in the tone-mapped images, and artifacts are present in severely exposed regions.
Liu et al. \cite{liu2020single} directly estimate and reverse the whole camera pipeline to generate HDR images. 
It sometimes struggles with generating detailed textures and produces artifacts in severely exposed regions, e.g., the over-exposed window frame and sky in the daylight.
With the {\AffineModule}, our model can preserve the image structure and generate the tone-mapped images with similar tones to the ground truth. 
It also produces fewer artifacts. 
More visual comparisons can be found in the supplementary material.

\begin{table}[]
\caption{\textbf{Ablation studies on the predicted LDR stack.}
{\AffineModuleHead} balances two distinct tasks: brightness adjustment and precise color tone generation while preserving the image structure. {\CycleTrainingHead} provides the model with extra information about changing EVs. 
Both designs improve LDR stack quality.
}
\centering
\resizebox{1.0\columnwidth}{!}{%
\begin{tabular}{lcccccc}
\toprule
Intensity transformation  & \multicolumn{2}{c}{-}  & \multicolumn{2}{c}{\checkmark}  & \multicolumn{2}{c}{\checkmark} \\ 
Cycle training & \multicolumn{2}{c}{-}  & \multicolumn{2}{c}{-}           & \multicolumn{2}{c}{\checkmark} \\ 
\midrule
\multirow{2}{*}{} & \multicolumn{2}{c}{PSNR} & \multicolumn{2}{c}{PSNR} & \multicolumn{2}{c}{PSNR} \\ \cline{2-7} 
   & m      & $\sigma$ & m      & $\sigma$ & m       & $\sigma$ \\ 
\midrule
EV+3 & 31.95  & 4.20   & 33.90  &  3.57  & \textbf{34.34}   & 3.46   \\
\midrule
EV+2 & 33.19  & 3.16   & 34.89  &  3.12  & \textbf{35.30}   & 3.08   \\ 
\midrule
EV+1 & 35.09  & 2.56   & 37.49  &  3.07  & \textbf{37.64}   & 2.96   \\ 
\midrule
EV-1 & 33.67  & 2.06   & 34.43  &  2.55  & \textbf{34.62}   & 3.47   \\ 
\midrule
EV-2 & 32.53  & 3.37   & 32.91  &  4.41  & \textbf{33.89}   & 4.34   \\
\midrule
EV-3 & 30.23  & 5.47   & 30.35  &  5.86  & \textbf{30.58}   & 5.32   \\ 
\bottomrule
\end{tabular}
}
\vspace{1mm}
\label{ABL_ThreeDesign_stack}
\vspace{-7mm}
\end{table}

\begin{table}[]
\caption{\textbf{Ablation studies on the reconstructed HDR images on the VDS dataset~\cite{Lee_2018}.} {\AffineModuleHead} and {\CycleTraining} enhance the quality of LDR stacks, and {\continuousStack} benefits the stack fusion process.}
\vspace{-3mm}
\centering
\resizebox{1.0\columnwidth}{!}{%
    \begin{tabular}{l|cccc}
    \toprule
    Intensity transformation    &   -   & \checkmark & \checkmark & \checkmark \\
    Continuous stack  &   -   &      -     & \checkmark & \checkmark \\
    Cycle training   &   -   &      -     &     -      & \checkmark \\ \midrule
    PSNR & 32.52 &    34.20   &    34.47   &   \textbf{34.67}    \\ \bottomrule
    \end{tabular}
}
\vspace{-3mm}
\label{ABL_ThreeDesign_HDR}
\end{table}

\subsection{Ablation Studies}
In the following, we validate three design contributions to improving the quality of the LDR stack and HDR images.

\begin{figure}[]
	\centering
	\includegraphics[width=\columnwidth]{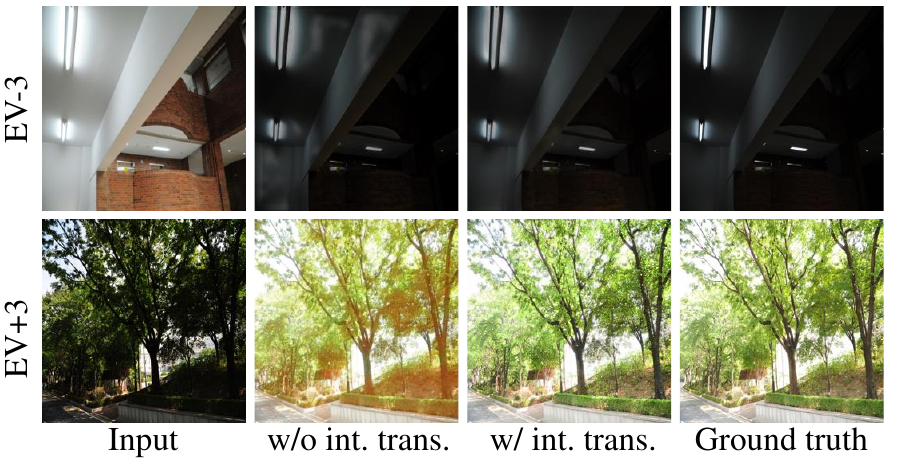}
	\vspace{-6mm}
	\caption{
        \textbf{Ablation on {\AffineModule}.} 
    With {\AffineModule}, CEVR can adjust LDR image intensity while preserving the image structure and color tone.
    }
	\label{ablation_affine}
	\vspace{-3mm}
\end{figure}

\topic{{\AffineModuleHead}.}
Learning to adjust image brightness while maintaining color tone accuracy and image structures can be challenging. 
The CEVR model, which directly outputs results from the U-net structure without using intensity transformation, can struggle to adjust brightness or produce inaccurate LDR images with artifacts, as shown in \cref{ablation_affine}. The intensity transformation module can restrict the model's capacity, producing LDR images with more accurate brightness, color tones, and structures. This design improves the quality of the LDR stack and HDR results, as demonstrated in \cref{ABL_ThreeDesign_stack} and \cref{ABL_ThreeDesign_HDR}.

\begin{figure}[]
	\centering
	\includegraphics[width=\columnwidth]{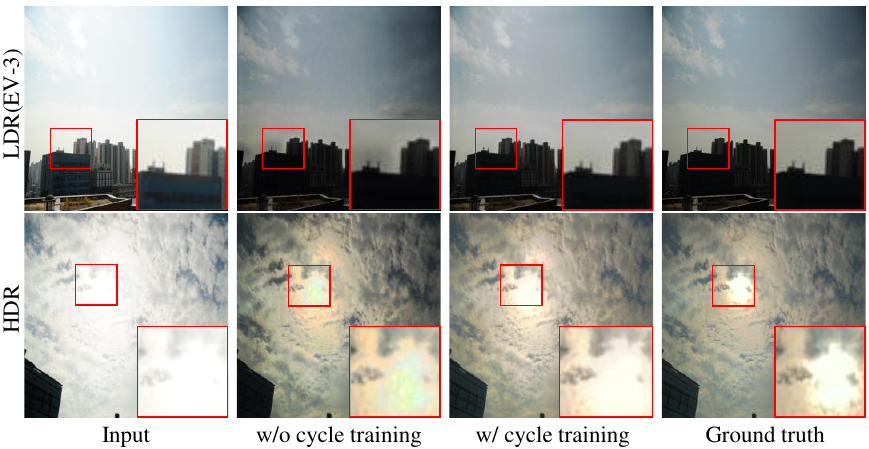}
	\vspace{-6mm}
    \caption{
        \textbf{Ablation on {\CycleTraining} for LDR and HDR images generation.} 
        With the {\CycleTraining}, the model captures the finer granularity of ``EV changing'' and generates more accurate and visually pleasing LDR and HDR images.
    }
	\label{ablation_cycle_LDR_HDR}
	\vspace{-3mm}
\end{figure}

\topic{\CycleTrainingHead.}
With \CycleTraining, the model can be supervised on continuous EV steps without using the corresponding ground truth. It can learn how to change the exposure value continuously, which improves the quality and reduces the artifacts of the estimated LDR images, which also leads to better HDR quality, as shown in \cref{ablation_cycle_LDR_HDR} ,\cref{ABL_ThreeDesign_stack} and  \cref{ABL_ThreeDesign_HDR}. To further demonstrate the effectiveness of \CycleTraining, we conducted the \emph{hold-out} experiment, excluding EV-1 and +1 LDR images during training. Then, we used the model to estimate EV-1 and +1 LDR images for each scene and evaluated the PSNR. The table shows that our model can generate better LDR images with unseen EVs when the cycle training strategy is adopted.

\begin{table}[h!]
    \caption{
    \textbf{Hold-out experiment.} Hold-out experiment excludes EV-1 and +1 LDR images during training. With \CycleTraining, the model generate better LDR images with unseen EVs (EV-1, +1).
    }
    \centering
    \resizebox{0.7\columnwidth}{!}{%
        \begin{tabular}{l|cc}
        \toprule
        Cycle training & $\times$ & $\checkmark$ \\
        \midrule
        EV-1              & 27.75              & \textbf{33.77}    \\ 
        \midrule
        EV+1              & 33.37              & \textbf{36.96}    \\
        \bottomrule
        \end{tabular}
    }
    \label{Hold-out_EXP}
\end{table}

\topic{{\continuousStackHead}.}
Debevec's method~\cite{debevec1997recovering} uses the LDR stack to recover response curves and reconstruct HDR images. A denser and continuous EV LDR stack helps produce an accurate inverse CRF, enhancing HDR quality. We compare two stack settings: ``predefined stack'' and ``continuous stack.'' The CEVR model estimates seven LDR images (EVs: -3, -2, -1, ...,  +3) for the predefined stack, which is the setting used in existing methods, while the continuous stack has 13 LDR images with various EVs (-3, -2.5, -2, ..., +3).~\cref{ABL_ThreeDesign_HDR} and \cref{teaser}(b)(c) show that the tone-mapped image from the continuous stack has superior quality and is more visually pleasing. We can further validate the effectiveness of the continuous stack by visualizing the CRF of both the predefined stack and the continuous stack. As shown in \cref{CRF}, the denser EV setting can help generate a smoother CRF  compared to the predefined EV setting. Additional analysis of the inverse CRF can be found in the supplementary material.

\subsection{Failure cases}
Although the proposed method performs favorably against other existing methods in quantitative and qualitative results, we do not explicitly design the module to address the over-exposed issue, which may make the CEVR model fail to generate reasonable content in large saturated regions, as shown in~\cref{failure_case}. It is a promising direction to take the emerging generative model designs, e.g.~\cite{zheng2022image, lugmayr2022repaint}, into account to address this issue.

\begin{figure}[]
	\centering
	\includegraphics[width=0.85\columnwidth]{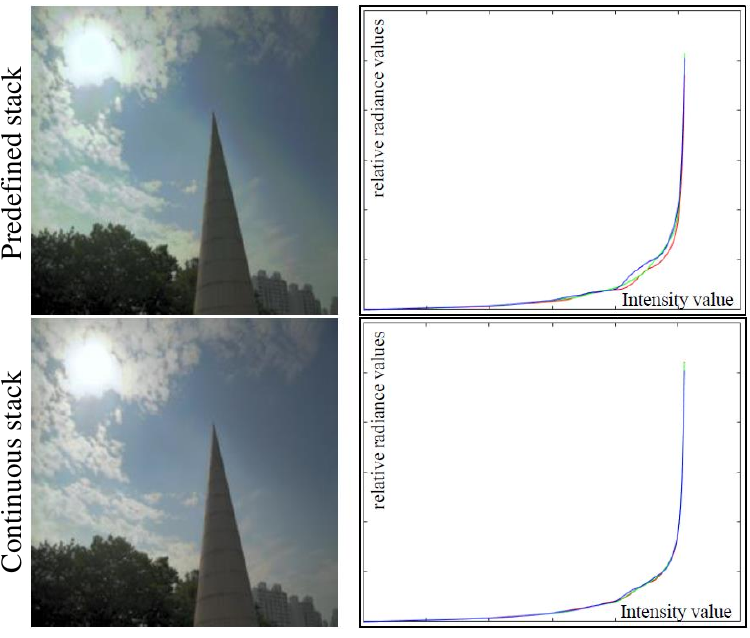}
	\caption{
        \textbf{Analysis of the estimated inverse CRF.}
        {\continuousStackHead}, adding additional LDR images with dense and continuous EVs, can help Debevec’s method~\cite{debevec1997recovering} generate more accurate inverse CRF and generate HDR images with better quality.
    }
	\label{CRF}
	    \vspace{-2mm}
\end{figure}

\begin{figure}[]
	\centering
	\includegraphics[width=\columnwidth]{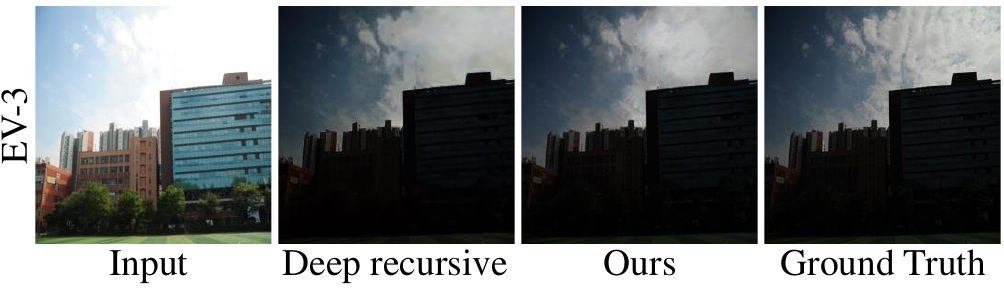}
	\vspace{-6mm}
    \caption{
            \textbf{Failure case.} 
            Existing methods and our proposed method cannot generate reasonable content as the sky region is severely over-exposed.
    }
	\label{failure_case}
	\vspace{-3mm}
\end{figure}
\section{Conclusion}
We introduce CEVR, a learning-based method that produces LDR EV stacks from continuous EV input. Our approach combines U-Net with implicit functions, and allows the network to generate LDR images with continuous EVs. We propose two strategies, including (1) cycle training for learning on continuous EV changes unseen in the training dataset and (2) continuous stack for improving LDR stack fusion using additional images with dense and continuous EVs. Our approach with the two strategies greatly enhances LDR stack quality and improves HDR image results, as demonstrated through extensive quantitative and qualitative evaluations on two benchmark datasets.

\topic{Acknowledgments.} This work was supported in part by National Science and Technology Council (NSTC) under grants 111-2628-E-A49-025-MY3, 112-2221-E-A49-090-MY3, 111-2634-F-002-023, 111-2634-F-006-012, 110-2221-E-A49-065-MY3 and 111-2634-F-A49-010. This work was funded in part by MediaTek.

{\small
\bibliographystyle{ieee_fullname}
\bibliography{egbib}
}

\clearpage
\newpage
\appendix

\section{Supplementary}
\label{sec:Supplementary}
We present additional results in the supplementary material to supplement our primary study. 
First, we create a
demo video to further demonstrate our CEVR model's flexibility in generating LDR images with continuous exposure values (EV), which is critical for our two major contributions, \textit{\continuousStack} and \textit{\CycleTraining}.
Second, we provide additional qualitative comparisons of LDR stacks and HDR images to demonstrate the effectiveness of our approach.
Finally, we demonstrate the effectiveness of {\continuousStack} by analyzing the estimated inverse camera response function (CRF) from Debevec’s method~\cite{debevec1997recovering}.

\subsection{Demo Video}
Existing methods~\cite{Lee_2018,lee2018deep}
mainly generate discrete EV LDR stacks and then fuse them to reconstruct HDR images. 
By contrast, our approach integrates the implicit neural representation into our model and makes it able to generate continuous EV LDR stacks. 
Based on this flexibility, we propose two main strategies, {\continuousStack} and {\CycleTraining}, to improve the quality of LDR stacks and HDR images. We also design the {\AffineModule} to further enhance the quality of estimated LDR images.
Our demo video (included in the supplementary files) shows the flexibility of our CEVR model in generating LDR images with continuous EVs and visualizes the effectiveness of our three main contributions (i.e., {\AffineModule},~{\CycleTraining}, and~{\continuousStack}).

\subsection{More Qualitative Comparisons}
Results presented in this section are generated with the model designs mentioned in the main manuscript ({\AffineModule}, {\continuousStack}, and {\CycleTraining}). 
Our model is trained on the training data in the VDS dataset~\cite{Lee_2018} and predicts the LDR images with different EVs from the testing data in the VDS and HDREye datasets~\cite{nemoto2015visual}.

\begin{figure*}[!ht]
	\centering
	    \includegraphics[width=1\linewidth]{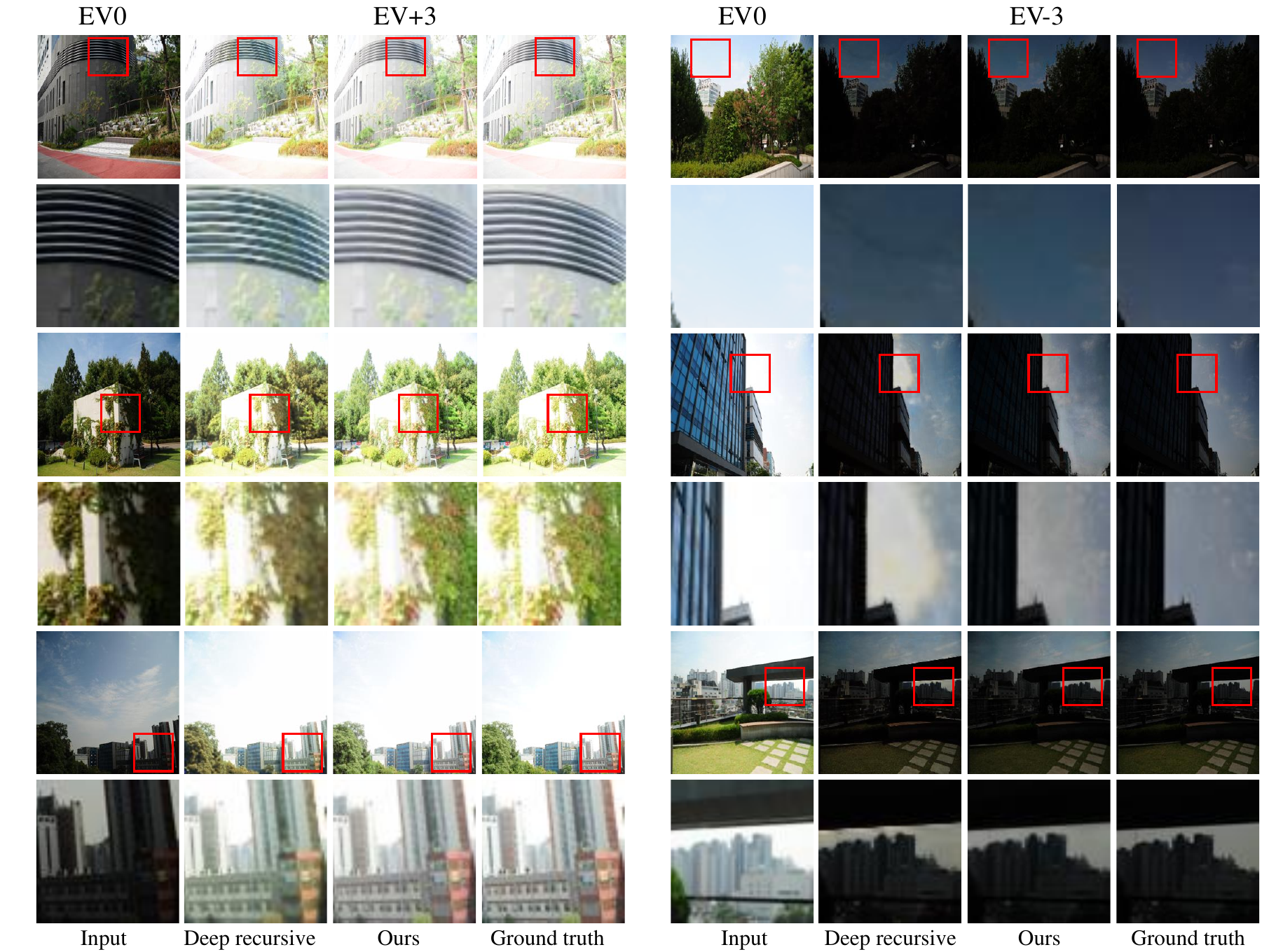}
	    \vspace{-5mm}
	\caption{
        \textbf{Extra qualitative comparisons of the LDR stack on the VDS dataset.} 
    With the proposed {\AffineModule} and {\CycleTraining}, our approach can generate high-quality estimated LDR images.
    }
	\label{extra_LDR_detail}
	    \vspace{-2mm}
\end{figure*}

\topic{Estimated LDR stack.}
We provide additional visual comparisons of the estimated LDR stack on the VDS dataset to verify the effectiveness of our approach compared to the existing approach, Deep recursive HDRI~\cite{lee2018deep}. 
As shown in~\cref{extra_LDR_detail}, our approach can predict LDR images with more accurate color tones while reducing artifacts.

\begin{figure*}[t!]
	\centering
	    \includegraphics[width=1\linewidth]{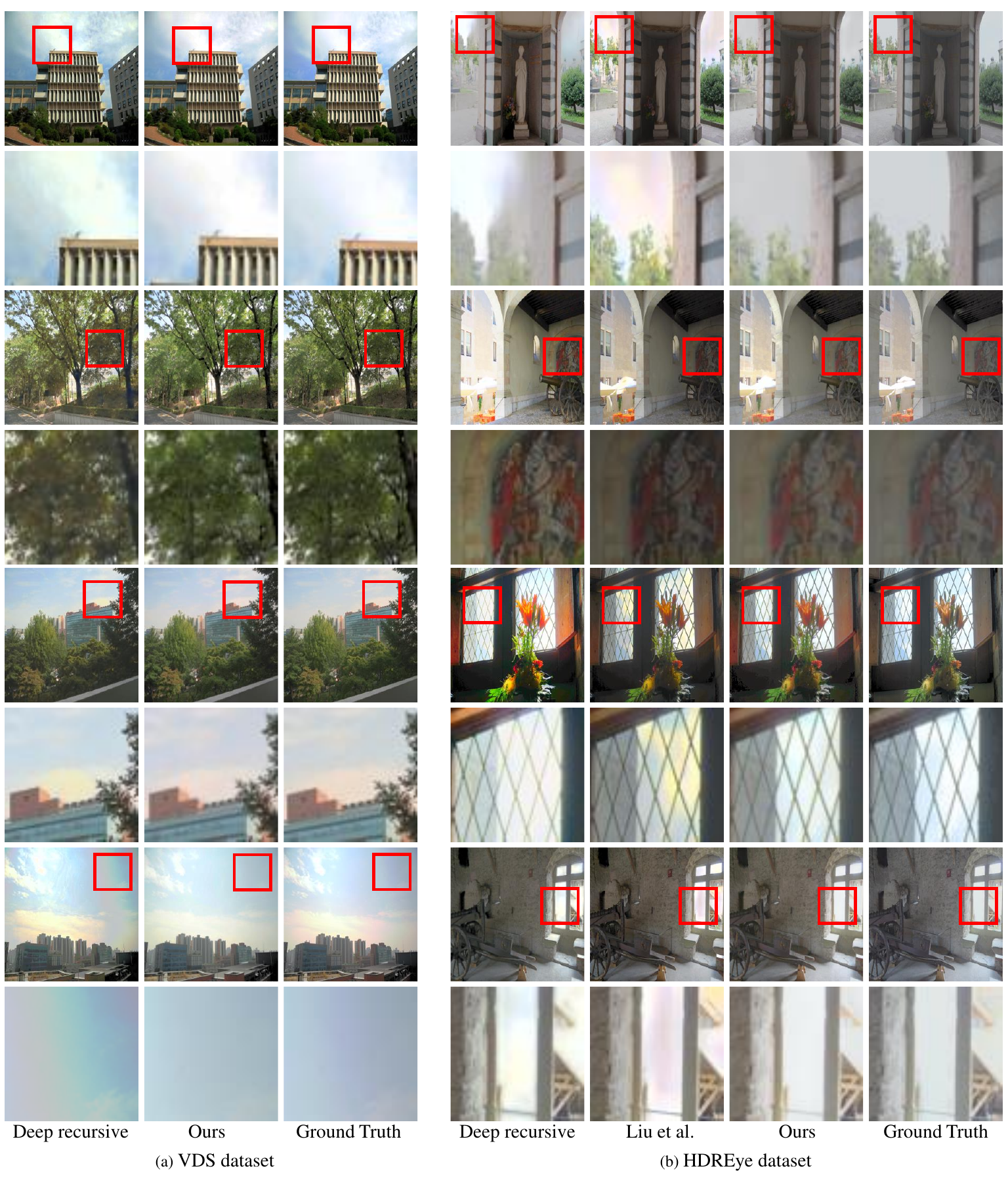}
	    \vspace{-5mm}
	\caption{
        \textbf{Extra qualitative comparisons of HDR images.}
        With the {\continuousStack}, our approach can generate more visually pleasing HDR images.
    }
	\label{extra_HDR_12}
	\vspace{-2mm}
\end{figure*}

\begin{figure*}[t!]
	\centering
	    \includegraphics[width=1\linewidth]{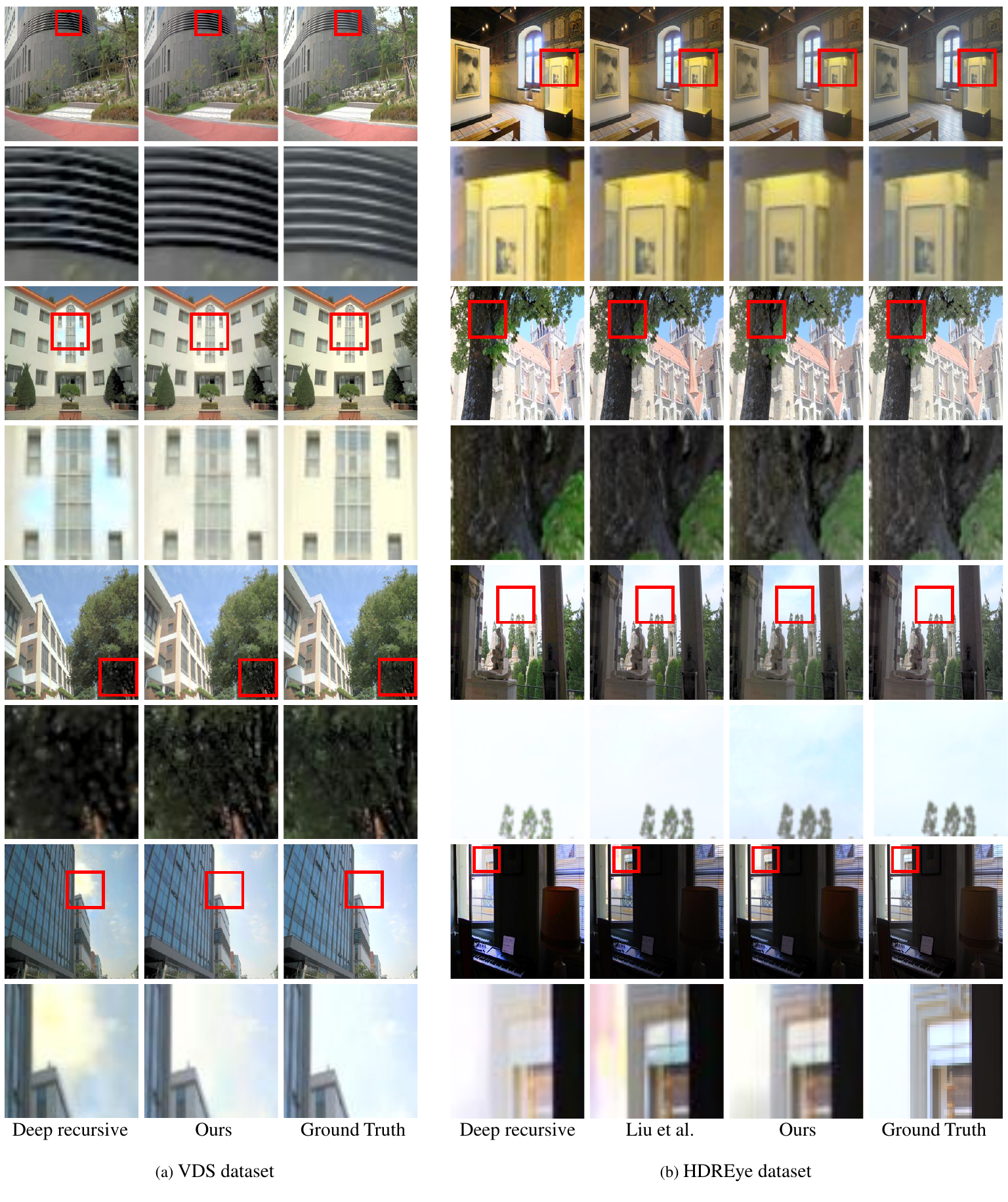}
	    \vspace{-5mm}
	\caption{
        \textbf{Extra qualitative comparisons of HDR images.}
        With the {\continuousStack}, our approach can generate more visually pleasing HDR images. In this figure, we conduct the same comparison as in~\cref{extra_HDR_12} but in different scenes.
    }
	\label{extra_HDR_34}
	    \vspace{-2mm}
\end{figure*}

\topic{Reconstructed HDR images.}
We showcase extra visual comparisons of reconstructed HDR images compared to the recent single-image HDR reconstruction methods, Deep recursive HDRI~\cite{lee2018deep} and Liu et al.~\cite{liu2020single}, on both the VDS and HDREye datasets.
As shown in~\cref{extra_HDR_12,extra_HDR_34}, our approach can generate tone-mapped images with better color tones and fewer artifacts.  

\subsection{Analysis of estimated inverse CRF}
Existing LDR stack-based methods, \eg ~\cite{Lee_2018,lee2018deep}, build the deep learning-generated LDR stack with predefined exposure values first, and Debevec’s method~\cite{debevec1997recovering} then generates the estimated inverse camera response function from the LDR stack and reconstructs the final HDR result. The inverse CRF, which should be \emph{monotonic} and \emph{smooth}, is used to transform the intensity value of LDR images into the relative radiance values of HDR images.

Based on the observation in~Fig. 2 in the main paper, we propose the CEVR model to generate an enriched and denser LDR stack. As shown in~\cref{CRF}, the estimated camera response function generated from the continuous stack setting is smoother than the one generated from the predefined stack setting. Due to the lack of ground truth camera response curves in both the VDS~\cite{Lee_2018} and HDREye~\cite{nemoto2015visual} datasets, we analyze the smoothness and monotonicity of the estimated CRF generated from two stack settings to evaluate the quality of estimated camera response curves.

\begin{figure*}[t!]
	\centering
	    \includegraphics[width=1\linewidth]{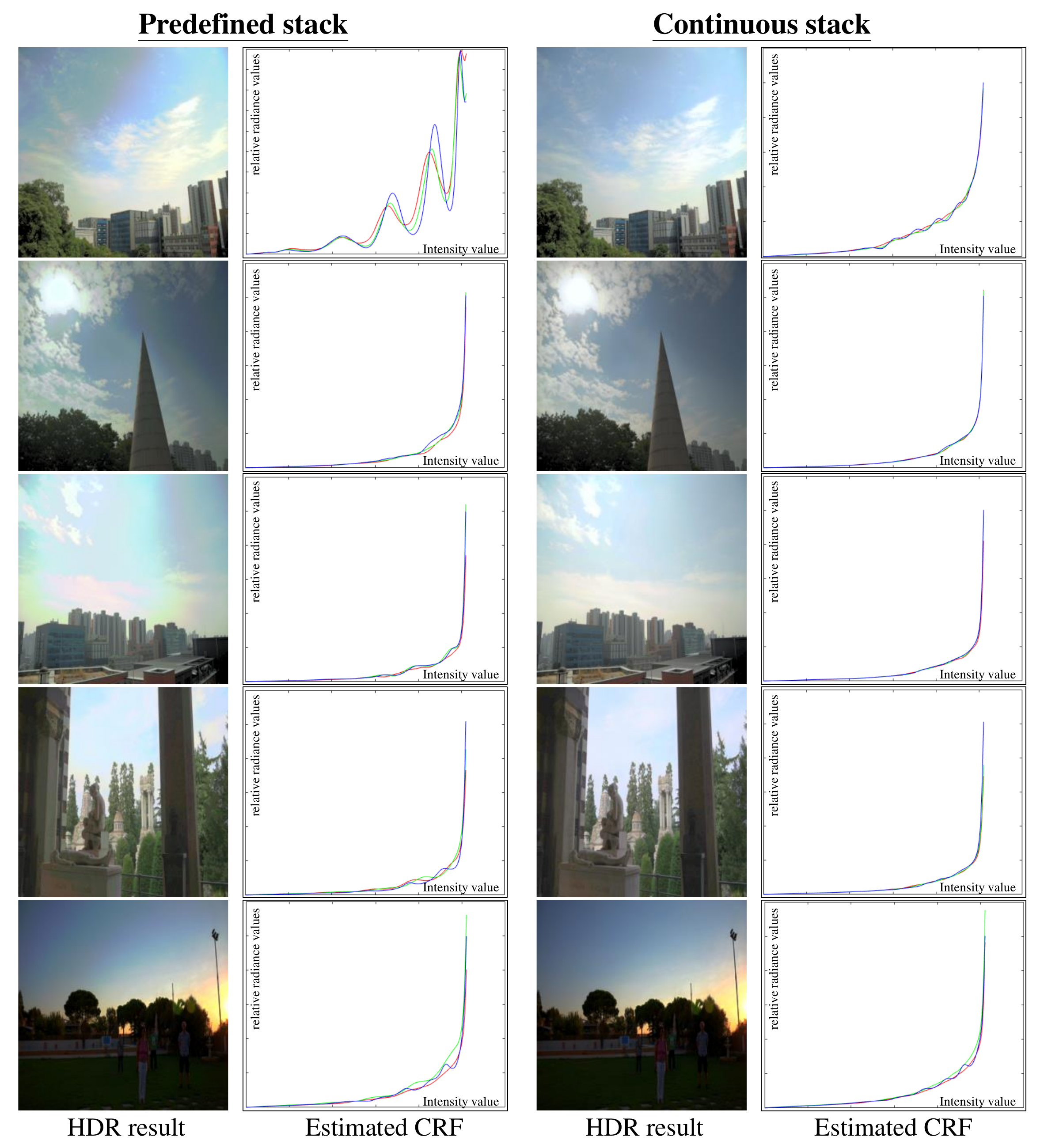}
	    \vspace{-5mm}
	\caption{
        \textbf{Analysis of the estimated inverse CRF.}
        {\continuousStackHead}, adding additional LDR images with dense and continuous EVs, can help Debevec’s method~\cite{debevec1997recovering} generate more accurate inverse CRF and generate HDR images with better quality.
    }
	\label{CRF}
	    \vspace{-2mm}
\end{figure*}

\end{document}